\documentclass[aps,prd,twocolumn,showpacs,showkeys,amsmath,amssymb]{revtex4}
\usepackage{graphicx}
\usepackage{bm}
\begin{document}

\title{$J^P=\frac{1}{2}^-$ Pentaquarks in Jaffe and Wilczek's Diquark Model}
\author{A. Zhang, Y.-R. Liu, P.-Z. Huang, W.-Z. Deng,
X.-L. Chen}
\affiliation{%
Department of Physics, Peking University, BEIJING 100871, CHINA}
\author{Shi-Lin Zhu}
\email{zhusl@th.phy.pku.edu.cn}
\affiliation{%
Department of Physics, Peking University, BEIJING 100871, CHINA}
\affiliation{%
COSPA, Department of Physics, National Taiwan University, Taipei
106, Taiwan, R.O.C.}

\date{\today}

\begin{abstract}

If Jaffe and Wilczek's diquark picture for $\Theta_5^+$ pentaquark
is correct, there should also exist a $SU_F$(3) pentaquark octet
and singlet with no orbital excitation between the diquark pair,
hence $J^P=\frac{1}{2}^-$. These states are lighter than the
$\Theta_5^+$ anti-decuplet and lie close to the orbitally excited
(L=1) three-quark states in the conventional quark model. We
calculate their masses and magnetic moments and discuss their
possible strong decays using the chiral Lagrangian formalism.
Among them two pentaquarks with nucleon quantum numbers may be
narrow. Selection rules of strong decays are derived. We propose
the experimental search of these nine additional
$J^P=\frac{1}{2}^-$ baryon states. Especially there are two
additional $J^P=\frac{1}{2}^-$ $\Lambda$ baryons around $\Lambda
(1405)$. We also discuss the interesting possibility of
interpreting $\Lambda (1405)$ as a pentaquark. The presence of
these additional states will provide strong support of the diquark
picture for the pentaquarks. If future experimental searches fail,
one has to re-evaluate the relevance of this picture for the
pentaquarks.
\end{abstract}


\maketitle

\pagenumbering{arabic}
\section{Introduction}\label{sec1}

Since LEPS announced the surprising discovery of the very narrow
$\Theta_5^+$ pentaquark ($uudd\bar s)$ around 1540 MeV last year
\cite{leps}, many other experimental groups have claimed the
observation of evidence of its existence
\cite{diana,clas,saphir,itep,clasnew,hermes,svd,cosy} while a few
groups reported negative results \cite{bes,hera-b}. Preliminary
experimental data indicate that $\Theta_5^+$ is an iso-scalar.
Later, NA49 Collaboration \cite{na49} reported a second narrow
pentaquark $\Xi_5^{--}$ $(ddss\bar u)$ at $1862$ MeV, to which
serious challenge is raised in Ref. \cite{doubt}. Very recently,
H1 Collaboration reported the discovery of the anti-charmed
pentaquark \cite{H1}.

One can use some textbook group theory to write down the wave
functions in the framework of quark model. Because of its low
mass, high orbital excitation with $L\ge 2$ is unlikely. Pauli
principle requires the totally anti-symmetric wave functions for
the four light quarks. Since the anti-quark is in the $[11]_C$
representation, the four quark color wave function is $[211]_C$.

With $L=0$, hence $P=-$, the 4q spatial wave function is
symmetric, i.e., $[4]_O$. Their $SU(6)_{FS}$ spin-flavor wave
function must be $[31]_{FS}^{210}$ which contains $[22]_F^6 \times
[31]_S^3$ after decomposition into $SU(3)_F\times SU(2)_S$
\cite{g1,g2,buccella}. Here 210, 6 etc is the dimension of the
representation. When combined with anti-quark, we get
$\left([33]_F^{10}+[21]_F^8\right)\times
\left([41]_S+[32]_S\right)$, which is nothing but $\left({\bar
{10}}_F+8_F\right)\times \left(({3\over 2})_S+({1\over
2})_S\right)$ in terms of more common notation. The total angular
momentum of the four quarks is one. The resulting exotic
anti-decuplet is always accompanied by a nearly degenerate octet.
Their angular momentum and parity is either $J^P={3\over 2}^-$ or
${1\over 2}^-$.

With $L=1$ and $P=+$, the four quark $SU(6)_{OFS}$
space-spin-flavor wave function must be $[31]_{OFS}$. Only this
representation can combine with $[211]_C$ color wave function to
ensure that the 4q total wave functions are anti-symmetric. The 4q
orbital wave function is $[31]_O$. There are several $SU(6)_{FS}$
wave functions $[4]_{FS}, [31]_{FS}, [22]_{FS}, [211]_{FS}$ which
allow the 4q total wave function to be anti-symmetric and lead to
the octet and exotic anti-decuplet \cite{g2,buccella}.

At present there are two outstanding pending issues: the
$\Theta_5^+$ parity and its narrow width. Two earlier lattice QCD
simulation favored negative parity for $\Theta_5^+$ \cite{lattice}
while a recent one advocates positive parity \cite{chiu}.
Theoretical papers can be roughly classified into two categories
according to $\Theta_5^+$ parity.

The parity of the anti-decuplet is positive in the chiral soliton
model \cite{diak,mp,ellis}. But the foundation of this framework
is questioned in Ref. \cite{cohen,princeton}. Several clustered
quark models were constructed to let the anti-decuplet carry
positive parity \cite{jaffe,lipkin,shuryak}. There are other
models favoring positive parity \cite{positive,carl,liuyx}.

On the other hand, QCD sum rule approach favors negative parity
for $\Theta_5^+$ \cite{zhu,qsr}. Recently in the framework of the
flux tube model $\Theta_5^+$ pentaquark is proposed to have a
extremely stable diamond structure with negative parity
\cite{song}. There are also many models supporting negative parity
\cite{zhang,carlson,wu}. Many schemes have been proposed to
determine the pentaquark parity experimentally \cite{proposal}.

All experiments indicate the $\Theta_5^+$ pentaquark is very
narrow. More stringent constraint on its decay width comes from
the reanalysis of the previous kaon nucleon scattering data, which
sets an upper bound of one or two MeV \cite{width}. Otherwise, the
$\Theta_5^+$ pentaquark should not have escaped detection. The
width of a conventional baryon 100 MeV above the threshold is
typically 100 MeV or bigger. Therefore, the extremely narrow width
is very puzzling.

Recently there appeared several interesting schemes for the narrow
width. Within the chiral soliton model, the coupling constants in
the leading order, next-leading order and next-next-leading order
large $N_c$ expansion cancel almost completely, which leads to a
narrow width \cite{ellis}. It is suggested that one of two nearly
degenerate pentaquarks can be arranged to decouple from the decay
modes after diagonalizing the mixing mass matrix via kaon nucleon
loop \cite{width2}. After constructing a special pentaquark wave
function with the color-orbital part being totally anti-symmetric,
the overlap amplitude between the final state and pentaquark is
suppressed significantly \cite{carlson}, which may also explain
the narrow width. Such a scheme is based on the mismatch of
initial and final state spin-flavor wave functions
\cite{maltman,close,buccella}. With the stable diamond structure
the system undergoes a special structural phase transition when
the $\Theta_5^+$ pentaquark decays into the planar kaon and
nucleon. The non-planar flux tubes were broken and new planar ones
are formed. Hence the decay width of the $\Theta_5^+$ pentaquark
should be small \cite{song}.

In this paper we will study the phenomenology of Jaffe and
Wilczek's diquark model for the pentaquarks. We note the problem
with with the identification of the ideally mixed positive parity
pentaquarks with the $N(1710)$ and $N(1440)$ is discussed
extensively in \cite{wrong}. If the diquark model is correct,
there should exist a $SU_F$(3) pentaquark octet and singlet with
no orbital excitation between the diquark pair and
$J^P=\frac{1}{2}^-$. These states are lighter than the
$\Theta_5^+$ anti-decuplet and lie close to the orbitally excited
(L=1) three-quark states in the conventional quark model.

Our paper is organized as follows: Section I is a brief review of
the field . In Section \ref{sec2} we use JW's model \cite{jaffe}
to calculate the masses and magnetic moments of the pentaquark
octet and singlet which arise from $\bf 3_F \otimes \bf
{\bar3_F}$, where the diquark-diquark system is in the flavor $\bf
3_F$ and the antiquark is in the flavor $\bf{\bar 3_F}$. Our
present pentaquark octet is in the mixed antisymmetric
representation, which is different from the mixed symmetric
pentaquark octet accompanying the anti-decuplet. In Section
\ref{sec3} we discuss strong decays of these states. In Section
\ref{sec4} we derive selection rules in the case of ideal mixing.
The final section is a short summary.

\section{Masses and magnetic moments of the pentaquark octet and singlet}\label{sec2}

Jaffe and wilczek \cite{jaffe} proposed that pentaquark states are
composed of two scalar diquarks and one anti-quark. Diquarks obey
Bose statistics. Each diquark is in the antisymmetric color
$\bf{\bar{3}}$ state. The spin wave function of the two quarks
within each scalar diquark is antisymmetric while the spatial part
is symmetric. Pauli principle requires the total wave function of
the two quarks in the diquark  be anti-symmetric. Thus the flavor
wave function of the two quarks in the diquark must be
antisymmetric, i.e, the diquark is in the flavor $\bf{\bar3_F}$
state. The diquark and antiquark flavor wave functions are listed
in Table \ref{tab1}.

\begin{table}[h]
\begin{center}
\begin{tabular}{c|c}\hline
($Y,I$, $I_3$)     &Flavor wave functions \\
\hline
($\frac{2}{3}$,0,0) & [ud],$\bar s$ \\
($-\frac{1}{3},\frac{1}{2},\frac{1}{2}$) & [su],$\bar d$ \\
($-\frac{1}{3},\frac{1}{2},-\frac{1}{2}$) & [ds],$\bar u$ \\
\hline
\end{tabular}
\end{center}
\caption{Diquark and antiquark flavor wave functions. Where $ Y ,
I$ and $I_3 $ are hypercharge, isospin and the third component of
isospin respectively. $[q_1 q_2]$ = $\frac{1}{\sqrt2}(q_1 q_2 -
q_2 q_1)$. }\label{tab1}
\end{table}

The color wave function of the two diquarks within the pentaquark
must be antisymmetric $\bf{3}_C$. In order to get an exotic
anti-decuplet, the two scalar diquarks combine into the symmetric
SU(3) $\bf{\bar{6}_F}$ : $[ud]^2$, $[ud][ds]_+$, $[su]^2$,
$[su][ds]_+$, $[ds]^2$, and $[ds][ud]_+$. Bose statistics demands
symmetric total wave function of the diquark-diquark system, which
leads to the antisymmetric spatial wave function with one orbital
excitation. The resulting anti-decuplet and octet pentaquarks have
$J^P={1\over 2}^+, {3\over 2}^+$.

We note that lighter pentaquarks can be formed if the two scalar
diquarks are in the antisymmetric $SU(3)_F$ $\bf 3$
representation: $[ud][su]_-$, $[ud][ds]_-$, and $[su][ds]_-$,
where
$[q_1q_2][q_3q_4]_-=\sqrt{\frac{1}{2}}([q_1q_2][q_3q_4]-[q_3q_4][q_1q_2])$.
No orbital excitation is needed to ensure the symmetric total wave
function of two diquarks since the spin-flavor-color part is
symmetric. The total angular momentum of these pentaquarks is
$\frac{1}{2}$ and the parity is negative. There is no accompanying
$J={3\over 2}$ multiplet. The two diquarks combine with the
antiquark to form a $SU(3)_F$ octet and singlet pentaquark
multiplet: ${\bar 3}_F \otimes 3_F = \bf{8}_F$ $\oplus$
$\bf{1}_F$. The flavor wave functions of the pentaquarks are
listed in Table \ref{tab2}. Similar mechanism has been proposed to
study heavy pentaquarks with negative parity and lighter mass than
$\Theta_{c,b}$ in \cite{wise,he}.

\begin{table}[h]
\begin{center}
\begin{tabular}{c|cc|c|c}\hline
&($Y,I$)               &   $I_3$       &Flavor wave functions&Masses (MeV) \\
\hline
$p_8$&(1,$\frac{1}{2}$)&$\frac{1}{2}$ &$[su][ud]_-\bar{s}$ &1460\\
$n_8$&         &-$\frac{1}{2}$ &$[ds][ud]_-\bar{s}$&1460 \\
$\Sigma_{8}^+$&(0,1)          &1 &$[su][ud]_-\bar{d}$ &1360\\
$\Sigma_{8}^0$&                &0  &$\frac{1}{\sqrt{2}}$($[su][ud]_-\bar{u}+[ds][ud]_-\bar{d}$)&1360 \\
$\Sigma_{8}^-$&                &-1          &$[ds][ud]_-\bar{u}$&1360 \\
$\Lambda_8$&(0,0)             & 0 &$\frac{[ud][su]_-\bar{u}+[ds][ud]_-\bar{d}-2[su][ds]_-\bar{s}}{\sqrt6}$&1533\\
$\Xi_{8}^0$&  (-1,$\frac{1}{2}$)  & $ \frac{1}{2}$               &$[ds][su]_-\bar{d}$&1520 \\
$\Xi_{8}^-$&                 &-$\frac{1}{2}$ &$[ds][su]_-\bar{u}$ &1520 \\
\hline
$\Lambda_1$&(0,0)&0                &$\frac{[ud][su]_-\bar{u}+[ds][ud]_-\bar{d}+[su][ds]_-\bar{s}}{\sqrt3}$&1447\\
\hline
\end{tabular}
\end{center}
\caption{Flavor wave functions and masses of the $\frac{1}{2}^-$
pentaquark octet and singlet. }\label{tab2}
\end{table}

According to JW's model \cite{jaffe}, the strange quark mass
explicitly breaks $SU(3)_F$ symmetry. The [ud] diquark is more
tightly bound than [us] and [ds]. The energy difference can be
related to the $\Sigma$-$\Lambda$ mass splitting. Thus, every
strange quark in the pentaquark contributes $\alpha \equiv \frac
{3}{4}(M_\Sigma - M_\Lambda) \approx 60$ MeV arising from [ud] and
[us], [ds] binding energy difference. The Hamiltonian in JW's
model reads
\begin{equation}
\ H_{s}= M_{0} +(n_{s}+n_{\bar{s}})m_{s}+n_{s} \alpha \label{eq1}
\end{equation}
where $M_{0}$ is the pentaquark mass in the $SU(3)_F$ symmetry
limit. The last two terms are from $SU(3)_F$ symmetry breaking
with $m_s\approx100$ MeV. $M_{0}$ has the form
\begin{equation}
M_{0}=2m_{di}+m_{\bar q}+\delta M_{l} \label{eq2}
\end{equation}
where $m_{di}$ is the [ud] diquark mass, $m_{\bar{q}}$ is the
anti-quark mass and $\delta M_{l}$ is the orbital excitation
energy. We follow Ref. \cite{jaffe} to use $m_{di}=420$ MeV,
$m_{\bar q}=360$ MeV and $\delta M_{l}=0$ for $l=0$ to get
$M_{0}=1200$ MeV. Thus we can use Eq.(\ref{eq1}) to compute masses
of the pentaquark octet and singlet. The numerical results are
collected in Table \ref{tab2}.

There are three mass relations among the nine pentaquarks. First
we get the Gell-Mann-Okubo relation for the pentaquark octet
\begin{equation}
2M_{N_8}+2M_{\Xi_8}=3M_{\Lambda_8}+M_{\Sigma_8},
\end{equation}
which is similar to that for the ground state octet. We also have
\begin{eqnarray}
M_{\Lambda_8}-M_{\Lambda_1}&=&M_{\Lambda_1}-M_{\Sigma_8}\nonumber\\
M_{\Lambda_8}-M_{N_8}&=&M_{\Xi_8}-M_{\Lambda_1}\;.
\end{eqnarray}

The pentaquark magnetic moment has the form \cite{huang}
\begin{equation}
\overrightarrow{\mu}=\sum\limits_{i}\overrightarrow{\mu_i}
=\sum\limits_{i}(g_{i}\overrightarrow{s_i}+\overrightarrow{l_i})
\mu_i,\label{eq3}
\end{equation}
where $\overrightarrow{s_i}$, $\overrightarrow{l_i}$ are the spin
and orbital momentum of the i-th constituent respectively. $g_i$
is the g-factor of the i-th constituent and $\mu_i$ is the
magneton of the i-th constituent. The spin of the scalar diquark
is zero. There is no orbital momentum. So the magnetic moment of
$J^P=\frac{1}{2}^-$ pentaquark $\overrightarrow{\mu}$ simply reads
 \begin{eqnarray}
\overrightarrow{\mu}&=& (g_1 \overrightarrow{0} +
\overrightarrow{0}) \mu_1 +(g_2 \overrightarrow{0} +
\overrightarrow{0}) \mu_2 +(g_3 \overrightarrow{\frac12} +
\overrightarrow{0}) \mu_3\nonumber\\
&=& g_3 \overrightarrow{\frac12} \mu_3 \; ,
\end{eqnarray}
where 1,2 denote the two scalar diquarks and 3 denotes the
anti-quark. It is clear that the pentaquark magnetic moment arises
from the anti-quark only. Finally we get
\begin{equation}\label{eq4}
\mu=\mu_{\bar q}=\frac{e_{\bar q}}{2m_{\bar
q}}=-\frac{e_{q}}{2m_{q}}
\end{equation}
where $e_{\bar q}$ is the charge of the antiquark and $m_{\bar q}$
is the mass of the antiquark. We present the expressions and
numerical results of octet and singlet pentaquark magnetic moments
in Table \ref{tab3}.

\begin{table}[h]
\begin{center}
\begin{tabular}{c|cc|c|c}\hline
&($Y,I$)               &   $I_3$       &Magnetic moments & Numerical results ($\mu_N$)\\
\hline
$p_8$&(1,$\frac{1}{2}$)&$\frac{1}{2}$&$\frac{e_0}{6m_s}$ &0.63\\
$n_8$&               &-$\frac{1}{2}$&$\frac{e_0}{6m_s}$ &0.63 \\
$\Sigma_{8}^+$&(0,1)          &1 &$\frac{e_0}{6m_d}$ &0.87\\
$\Sigma_{8}^0$&                &0                &$\frac{1}{6}(-\frac{e_0}{m_u}+\frac{e_0}{2m_d})$&-0.43\\
$\Sigma_{8}^-$&                &-1          &$-\frac{e_0}{3m_u}$&-1.74\\
$\Lambda_8$  &(0,0)& 0 &$\frac{1}{18}$(-$\frac{e_0}{m_u}$+$\frac{e_0}{2m_d}$+$\frac{2e_0}{m_s}$)&0.27\\
$\Xi_{8}^0$& (-1,$\frac{1}{2}$)  & $ \frac{1}{2}$               &$\frac{e_0}{6m_d}$ &0.87\\
$\Xi_{8}^-$&               &-$\frac{1}{2}$ &-$\frac{e_0}{3m_u}$ &-1.74\\
\hline
$\Lambda_1$&(0,0)&0                &$\frac{1}{9}(-\frac{e_0}{m_u}+\frac{e_0}{2m_d}+\frac{e_0}{2m_s})$&-0.08\\
\hline
\end{tabular}
\end{center}
\caption{Expressions and numerical results of the  magnetic
moments of the pentaquark octet and singlet, where $e_0$ is the
charge unit. } \label{tab3}
\end{table}

There exist several magnetic moment relations.
\begin{eqnarray}
\mu_{\Lambda_8}-\mu_{\Sigma_8^0}&=&2(\mu_{n_8}-\mu_{\Lambda_8})
=2(\mu_{\Lambda_8}-\mu_{\Lambda_1})\nonumber\\
\mu_{\Sigma_8^+}-\mu_{\Sigma_8^0}&=&\mu_{\Sigma_8^0}-\mu_{\Sigma_8^-}\nonumber\\
\mu_{n_8}-\mu_{\Lambda_1}&=&2(\mu_{\Lambda_1}-\mu_{\Sigma_8^0})\nonumber\\
\mu_{p_8}&=&\mu_{n_8}\nonumber\\
\mu_{\Xi_8^0}&=&\mu_{\Sigma_8^+}\nonumber\\
\mu_{\Xi_8^-}&=&\mu_{\Sigma_8^-}\;.
\end{eqnarray}
We note only the second one is similar to the Coleman-Glashow
relations for nucleon octet \cite{coleman}.

\section{Pentaquark Chiral Lagrangian, Strong Decays And Selection Rules}\label{sec3}

In the case of pentaquark decays, if symmetry and kinematics
allow, the most efficient decay mechanism is for the four quarks
and anti-quark to regroup with each other into a three-quark
baryon and a meson. This is in contrast to the $^3P_0$ decay
models for the ordinary hadrons. This regrouping is coined as the
"fall-apart" mechanism in Refs.
\cite{maltman,close,carlson,buccella}.

In the following we write down the interaction chiral Lagrangian
using $SU(3)_F$ symmetry. We denote a quark and anti-quark by
$q^i, {\bar q}_j$ where $i, j$ are the $SU(3)_F$ flavor indices.
Note that the flavor wave function of the $J^P={1\over 2}^-$ octet
and singlet pentaquark arise from
\begin{equation}
A_{[ij]}\otimes\bar{q}_k=S\oplus O_{[ij,k]},
\end{equation}
where the indices $ij$ are antisymmetric, $A_{[ij]}$ is the
$\bf{3}_F$ diquark pair. $S$ is the pentaquark singlet whose
indices are contracted completely. $O_{[ij,k]}$ is the octet
representation. The index $k$ represents the antiquark which
contracts with one of the meson index.

In Ref. \cite{lee} the chiral Lagrangian is built to discuss the
decay modes of the anti-decuplet and octet with positive parity.
The authors pointed out that keeping explicit track of the flavor
indices of the two diquarks minimize the independent coupling
constants and lead to some selection rules.

For the interaction of the $J^P={1\over 2}^-$ pentaquark octet
$P$, nucleon octet $B$ and pseudoscalar meson octet $M$, we have
\begin{equation}
{\cal{L}}_8=g_8\epsilon_{ilm}\bar{O}^{[ij,k]}B^l_jM^m_k+H.c.,
\end{equation}
where $O_{[ij,k]}=\epsilon_{ljk}P^l_i-\epsilon_{lik}P^l_j$. The
explicit form of the matrix $B^i_j$, $M^i_j$ and $P^i_j$ is
\begin{eqnarray}
(P_j^i)&=& \left(\begin{array}{ccc}
\frac{\Sigma_8^0}{\sqrt{2}}+\frac{\Lambda_8}{\sqrt{6}}&\Sigma_8^+&p_8\\
 \Sigma_8^-&-\frac{\Sigma_8^0}{\sqrt{2}}+\frac{\Lambda_8}{\sqrt{6}}&n_8\\
\Xi_8^-&\Xi_8^0&-\frac{2\Lambda_8}{\sqrt{6}} \end{array}\right),\\
(B_j^i)&=&\left(\begin{array}{ccc}
\frac{\Sigma^0}{\sqrt{2}}+\frac{\Lambda}{\sqrt{6}}&\Sigma^+&p\\
 \Sigma^-&-\frac{\Sigma^0}{\sqrt{2}}+\frac{\Lambda}{\sqrt{6}}&n\\
\Xi^-&\Xi^0&-\frac{2\Lambda}{\sqrt{6}} \end{array}\right),\\
(M_j^i)&=&\left(\begin{array}{ccc}
\frac{\pi^0}{\sqrt{2}}+\frac{\eta_0}{\sqrt{6}}&\pi^+&K^+\\
 \pi^-&-\frac{\pi^0}{\sqrt{2}}+\frac{\eta_0}{\sqrt{6}}&K^0\\
K^-&\bar{K}^0&-\frac{2\eta_0}{\sqrt{6}} \end{array}\right).
\end{eqnarray}
We present the Clebsch-Gordan coefficient of each interaction term
in Table \ref{tab5}.

\begin{table}[h]
\begin{center}
\begin{tabular}{cc|cc|cc|cc}\hline
$\Xi^-_8$             &  &   $\Xi^0_8$&       &$p_8$ && $n_8$&\\
\hline

$\Xi^-\pi^0$ & $\frac{1}{\sqrt{2}}$ & $\Xi^-\pi^+$ & 1 & $\Sigma^0 K^+$ &  $\frac{1}{\sqrt{2}}$  &$\Sigma^- K^+$ &  1\\
$ \Xi^0\pi^-$ & 1 & $\Xi^0\pi^0$ &-$\frac{1}{\sqrt{2}}$  & $\Sigma^+ K^0$ &  1 &$\Sigma^0 K^0$ &-$\frac{1}{\sqrt{2}}$\\
$\Xi^-\eta_0$ & $\frac{1}{\sqrt{6}}$ & $\Xi^0\eta_0$ &$\frac{1}{\sqrt{6}}$  & $p\eta_0$ &-$\frac{2}{\sqrt{6}}$ &$n\eta_0$ &-$\frac{2}{\sqrt{6}}$\\
$\Lambda K^-$ &-$\frac{2}{\sqrt{6}}$ & $\Lambda\bar{K^0}$ &-$\frac{2}{\sqrt{6}}$  &$\Lambda K^+$ & $\frac{1}{\sqrt{6}}$ &$\Lambda K^+$ & $\frac{1}{\sqrt{6}}$\\
 \hline
$\Sigma^0_8$             &  &   $\Sigma^+_8$&       &$\Sigma^-_8$ && $\Lambda_8$&\\
\hline
$\Sigma^+ \pi^-$&$\frac{1}{\sqrt{2}}$&$\Sigma^+\pi^0$&-$\frac{1}{\sqrt{2}}$&$\Sigma^-\pi^0$&$\frac{1}{\sqrt{2}}$ &$\Sigma^+\pi^-$&$\frac{1}{\sqrt{6}}$\\
$\Sigma^- \pi^+$&-$\frac{1}{\sqrt{2}}$&$\Sigma^0\pi^+$&$\frac{1}{\sqrt{2}}$&$\Sigma^0\pi^-$&-$\frac{1}{\sqrt{2}}$&$\Sigma^-\pi^+$&$\frac{1}{\sqrt{6}}$\\
$\Sigma^0 \eta_0$&$\frac{1}{\sqrt{6}}$&$\Sigma^+\eta_0$&$\frac{1}{\sqrt{6}}$&$\Sigma^-\eta_0$&$\frac{1}{\sqrt{6}}$ &$\Sigma^0\pi^0$&$\frac{1}{\sqrt{6}}$\\
$p K^-$&$\frac{1}{\sqrt{2}}$&$p \bar{K^0}$& 1 &$n K^-$& 1 &$p K^-$&$\frac{1}{\sqrt{6}}$\\
$n \bar{K^0}$&-$\frac{1}{\sqrt{2}}$&$\Lambda\pi^+$& $\frac{1}{\sqrt{6}}$ &$\Lambda\pi^-$& $\frac{1}{\sqrt{6}}$&$n \bar{K^0}$&$\frac{1}{\sqrt{6}}$\\
$\Lambda\pi^+$& $\frac{1}{\sqrt{6}}$ & & &&& $\Xi^- K^+$&-$\frac{2}{\sqrt{6}}$\\
 &   & & & & & $\Xi^0 K^0$&-$\frac{2}{\sqrt{6}}$\\
 &   & & & & & $\Lambda \eta_0$&-$\frac{1}{\sqrt{6}}$\\
 \hline
\end{tabular}
\end{center}
\caption{Coupling of the $J^P=\frac12^-$ pentaquark octet with the
usual baryon octet and the pseudoscalar meson octet. The universal
coupling constant $g_8$ is omitted.  } \label{tab5}
\end{table}

The pentaquark octet can also couple with usual baryon octet and
meson singlet $\eta_1$.
\begin{eqnarray}\label{eqn9}
{\cal{L}}_1&=&g_1\bar{P}^j_iB^i_j\eta_1+H.c.\nonumber\\
&=&g_1(\bar{N}_8N+\bar{\Sigma}_8\Sigma+\bar{\Xi}_8\Xi+\bar{\Lambda}_8\Lambda)\eta_1+H.c,
\end{eqnarray}
where
\begin{eqnarray}
 N = \left( \begin{array}{c} p \\ n \end{array} \right),
\quad \Xi = \left( \begin{array}{c} \Xi^0 \\ \Xi^- \end{array}
\right), \quad \Sigma = \left( \begin{array}{c} \Sigma^+ \\
\Sigma^0 \\ \Sigma^-
\end{array} \right).
\end{eqnarray}

The interaction among pentaquark singlet $\Lambda_1$, normal
baryon octet $B$ and meson octet $M$ is
\begin{eqnarray}
{\cal{L}}_1^\prime&=&G_1\bar{\Lambda}_1B^i_jM^j_i+H.c.\nonumber\\
&=&G_1\bar{\Lambda}_1(K_c N +\pi\Sigma+K\Xi+\eta_0\Lambda)+H.c.,
\end{eqnarray}
where
\begin{eqnarray}
 \pi = \left(\pi^-,   \pi^0,  \pi^+  \right),
 K = \left(  K^0, K^+ \right),  K_c = \left( K^-
,\bar{K}^0 \right).
\end{eqnarray}

Since the parity of these pentaquarks is negative, they will decay
via S-wave if kinematics allows. With the mass values in Table
\ref{tab2}, it is easy to find out which decay process will occur.

According to our mass estimate, only $p_8, n_8$ are below the
threshold of the listed decay modes in Table \ref{tab5}. At first
sight, there is no strong decay modes for them. They should be
stable particles.

However, there are multiple-pion decays modes which violate the
"fall-apart" mechanism, such as S-wave $p_8\to N\pi$, P-wave $p_8
\to N \pi \pi$ and S-wave $p_8\to N \pi\pi\pi$ where $N$ is either
a proton or neutron.

Another possibility is the isospin violating strong decay mode
$p_8\to p\eta_0\to p\pi$. The virtual intermediate state $p\eta_0$
helps this process happen. The first step satisfies the
"fall-apart" mechanism. Then the virtual $\eta_0$ turns into a
real pion through isospin violating effects. All these processes
contribute to the decay width of $p_8, n_8$. However both $p_8$
and $n_8$ should still be narrow resonances.

\section{Additional Selection Rules In The Ideal mixing Case For The $I=0$ Sector}\label{sec4}

For the $I=0$ channel, physical states are the mixture of octet
and singlet states. For example, the physical $\eta, \eta'$ are
the mixture of $\eta_0$ and $\eta_1$ where $\eta_0$ is the pure
octet member and $\eta_1$ is the pure singlet. In the following we
will let the mixing angle deviate from the physical value.
\begin{eqnarray}
\eta&=&\eta_0\cos\theta-\eta_1\sin\theta \nonumber\\
\eta^\prime&=&\eta_0\sin\theta+\eta_1\cos\theta \; .
\end{eqnarray}
From the above we have
\begin{eqnarray}
\eta_0&=&\eta^\prime\sin\theta+\eta\cos\theta \nonumber\\
\eta_1&=&\eta^\prime\cos\theta-\eta\sin\theta  \;.
\end{eqnarray}

The mixing of $\Lambda_8$ and $\Lambda_1$ is defined as
\begin{eqnarray}
\Lambda_n&=&\Lambda_8\cos\varphi-\Lambda_1\sin\varphi \nonumber\\
\Lambda_s&=&\Lambda_8\sin\varphi+\Lambda_1\cos\varphi.
\end{eqnarray}

Now the interaction terms involving $I=0$ states are
\begin{eqnarray}\label{eq15}
&&{\cal{L}}_{mixing}=g_8\{(\bar\Sigma_8\Sigma+\bar\Xi_8\Xi)\eta^\prime(\frac{1}{\sqrt6}\sin\theta+a\cos\theta)\nonumber\\
&&+(\bar\Sigma_8\Sigma+\bar\Xi_8\Xi)\eta(\frac{1}{\sqrt6}\cos\theta-a\sin\theta)\nonumber\\
&&+\bar N_8N\eta^\prime(-\frac{2}{\sqrt6}\sin\theta+a\cos\theta)\nonumber\\
&&+\bar N_8N\eta(-\frac{2}{\sqrt6}\cos\theta-a\sin\theta)\nonumber\\
&&+\bar\Lambda_s(\pi\Sigma+K_cN)(\frac{1}{\sqrt6}\sin\varphi+b\cos\varphi)\nonumber\\
&&+\bar\Lambda_n(\pi\Sigma+K_cN)(\frac{1}{\sqrt6}\cos\varphi-b\sin\varphi)\nonumber\\
&&+\bar\Lambda_sK\Xi(-\frac{2}{\sqrt6}\sin\varphi+b\cos\varphi)\nonumber\\
&&+\bar\Lambda_nK\Xi(-\frac{2}{\sqrt6}\cos\varphi-b\sin\varphi)\nonumber\\
&&+\bar\Lambda_s\Lambda\eta\prime(-\frac{1}{\sqrt6}\sin\varphi\sin\theta+a\sin\varphi\cos\theta+b\cos\varphi\sin\theta)\nonumber\\
&&+\bar\Lambda_s\Lambda\eta(-\frac{1}{\sqrt6}\sin\varphi\cos\theta-a\sin\varphi\sin\theta+b\cos\varphi\cos\theta)\nonumber\\
&&+\bar\Lambda_n\Lambda\eta\prime(-\frac{1}{\sqrt6}\cos\varphi\sin\theta+a\cos\varphi\cos\theta-b\sin\varphi\sin\theta)\nonumber\\
&&+\bar\Lambda_n\Lambda\eta(-\frac{1}{\sqrt6}\cos\varphi\cos\theta-a\cos\varphi\sin\theta-b\sin\varphi\cos\theta)\}\nonumber\\
&&+H.c. \;.
\end{eqnarray}
where $a={g_1\over g_8}, b={G_1\over g_8}$.

In the extreme case of ideal mixing, i.e.,
$\tan\theta=\tan\varphi=-\sqrt{2}$, we have
\begin{eqnarray}
\Lambda_s&=&[su][ds]_-\bar s\nonumber\\
\Lambda_n&=&\frac{1}{\sqrt2}([ud][su]_-\bar u+[ds][ud]_-\bar d)\nonumber\\
\eta^\prime&=&s\bar s\nonumber\\
\eta&=&\frac{1}{\sqrt2}(u\bar u+d\bar d)\;.
\end{eqnarray}
The so-called "fall-apart" mechanism requires that there is no
annihilation or creation of quark pairs when pentaquarks decay.
Hence the coefficient of the fourth and eighth terms must vanish
in the limit of ideal mixing. In this way, we get
\begin{equation}
a=b=\frac{1}{\sqrt{3}}\; .
\end{equation}
The three coupling constants $g_1, G_1, g_8$ are related to each
other in this limit.

Now Eq. (\ref{eq15}) has a simple form
\begin{eqnarray}\nonumber
&&{\cal{L}}_{mixing}=\frac{1}{\sqrt2}(\bar\Sigma_8\Sigma+\bar\Xi_8\Xi)\eta+\bar
N_8N\eta^\prime\\ \nonumber
&&+\frac{1}{\sqrt2}\bar\Lambda_n(\pi\Sigma+K_cN)\nonumber
+\bar\Lambda_sK\Xi\\
&&-\frac{2}{\sqrt6}\bar\Lambda_s\Lambda\eta^\prime
+\frac{1}{\sqrt6}\bar\Lambda_n\Lambda\eta+(H.c.)\;.
\end{eqnarray}
The decay modes of $\Lambda_s$ and $\Lambda_n$ are
\begin{eqnarray}
\Lambda_s &\longrightarrow&
K\Xi-\frac{2}{\sqrt6}\Lambda\eta^\prime\\
\Lambda_n&\longrightarrow&\frac{1}{\sqrt2}(\pi\Sigma+K_cN)+\frac{1}{\sqrt6}
\Lambda\eta\;.
\end{eqnarray}
The above relation is the selection rule from the "fall-apart"
mechanism in the ideal mixing limit.

It's very interesting to note that the only dynamically allowed
two decay modes of $\Lambda_s$ are kinematically forbidden since
$\Lambda_s$ is below the threshold. Therefore, $\Lambda_s$ will
not decay via strong interaction. It is a long-lived stable
particle in the ideal mixing case. For $\Lambda_n$ the only both
dynamically and kinematically allowed decay mode is $\pi\Sigma$.
Unfortunately, the physical $\eta$ and $\eta'$ are not ideally
mixed. So the results obtained in this section may be different
from the realistic case, therefore not very useful.

\section{Discussion}\label{sec5}

We have shown that there exist an octet and singlet pentaquark
multiplet with $J^P={1\over 2}^-$ in the framework of Jaffe and
Wilczek's diquark model. We have calculated their masses and
magnetic moments. Several interesting mass and magnetic moment
relations are derived. We have also constructed the chiral
Lagrangian for these pentaquarks. Possible strong decay modes are
discussed. We have derived selection rules based on the
"fall-apart" decay mechanism. In this limit there exists a
long-lived stable $J^P={1\over 2}^-$ $\Lambda_s$ pentaquark which
will not decay via strong interaction.

Because there is no orbital excitation within these nine
pentaquarks, their masses are lower than the anti-decuplet and the
accompanying octet with positive parity. Their masses range
between 1360 MeV and 1540 MeV according to our calculation using
the same mass formula in \cite{jaffe}. These states are close to
the L=1 orbital excitations of the nucleon octet. The mixing
between the pentaquark states and orbital excitations is expected
to be small since their spatial wave functions are very different.

According to our calculation, two of the $J^P={1\over 2}^-$ octet
pentaquark members $p_8, n_8$ lie 22 MeV below the $p\eta_0$
threshold and 228 MeV below the $\Sigma K$ threshold. The
"fall-apart" decay mechanism forbids $p_8$ to decay into one
nucleon and one pion. Although their interaction is of S-wave,
lack of phase space forbids the strong decays $p_8\to p\eta_0,
\Sigma K$ to happen.

For $p_8$, the only kinematically strong decays are S-wave $p_8\to
p\pi$, P-wave $p_8 \to N \pi \pi$ and S-wave $p_8\to N \pi\pi\pi$
decays where $N$ is either a proton or neutron. All these decay
modes involve the anihilation of a strange quark pair and violate
the the "fall-apart" mechanism. Hence the width is expected to be
small. Both $p_8$ and $n_8$ may be narrow resonances.

It is interesting to note there are three negative-parity
$\Lambda$ particles within the range between 1400 MeV and 1540 MeV
if Jaffe and Wilczek's diquark model is correct. One of them is
the well established $\Lambda (1405)$. $\Lambda (1405)$ is only 30
MeV below kaon and nucleon threshold. Some people postulated it to
be a kaon nucleon molecule \cite{pdg}. We propose that there is
another intriguing possibility of interpreting $\Lambda (1405)$ as
the candidate of $J^P={1\over 2}^-$ pentaquark. The other
$J^P={1\over 2}^-$ pentaquark and the corresponding L=1 singlet
$\Lambda$ particle may have escaped detection so far.

The discovery of nine additional negative-parity baryons in this
mass range will be strong evidence supporting the diquark model.
On the other hand, if future experimental searches fail to find
any evidence of these additional states with negative parity, one
has to re-evaluate the relevance of the diquark picture for the
pentaquarks.

S.L.Z. thanks Prof W.-Y. P. Hwang and COSPA center at National
Taiwan University for the warm hospitality. S.L.Z. thanks Prof T D
Cohen for informing us of his similar ideas of negative-parity
octet in his recent talk \cite{talk} after our paper appeared in
the e-print archive. This project was supported by the National
Natural Science Foundation of China under Grant 10375003, Ministry
of Education of China, FANEDD and SRF for ROCS, SEM.


\end{document}